\documentclass[aps,prl,twocolumn,superscriptaddress,longbibliography,nofootinbib]{revtex4-2}

\usepackage{float}

\usepackage[T1]{fontenc}
\usepackage{lmodern}
\usepackage{amsmath,amssymb,bm,graphicx}
\usepackage{microtype}
\usepackage[colorlinks=true,citecolor=blue,urlcolor=blue,linkcolor=blue]{hyperref}
\usepackage{url}
\usepackage{hyphenat}
\graphicspath{{./}}
\usepackage{xcolor}

\renewcommand{\doi}[1]{\href{https://doi.org/#1}{doi:#1}}

\begin{document}

\title{How do late-time photons ``record'' primordial curvature perturbations?}
\author{Tomotaka Kuroda}
\email{kuroda.t.ad@m.titech.ac.jp}
\affiliation{Department of Physics, Institute of Science Tokyo, Tokyo 152-8551, Japan}
\affiliation{Cosmology, Gravity, and Astroparticle Physics Group, Center for Theoretical Physics of the Universe, Institute for Basic Science (IBS), Daejeon 34126, Korea}

\author{Fumiya Sano}
\email{sanof.cosmo@gmail.com}
\affiliation{Graduate School of Arts and Sciences, The University of Tokyo, Tokyo 153-8902, Japan}

\author{Yusuke Yamada}
\email{yamada@ibs.re.kr}
\affiliation{Cosmology, Gravity, and Astroparticle Physics Group, Center for Theoretical Physics of the Universe, Institute for Basic Science (IBS), Daejeon 34126, Korea}

\begin{abstract}
Inflationary cosmology faces a fundamental gap between primordial quantum fluctuations and the classical cosmic history inferred from observations. Although the primordial quantum state is a superposition of field configurations, standard cosmology assumes one definite realization of curvature perturbations and uses classical perturbation theory to describe the resulting gravitational potentials, CMB anisotropies, and large-scale structure. Squeezing and decoherence of primordial fluctuations alone do not explain why this one-realization description is physically meaningful. We address this issue through Cosmological Classical Branches (CCBs), using the quantum-information-theoretic notion of physical records. In CCBs, distinct late-time matter components carry compatible records of the same primordial configuration. A record in one component makes that configuration readable, while a disjoint compatible record suppresses interference accessible to operations on the first. Each branch thus admits a common c-number primordial label at finite resolution, while the global quantum superposition persists. This Letter summarizes the framework developed in Ref.~\cite{Kuroda2026CCB}.
\end{abstract}

\maketitle

\emph{Introduction.---}
One of the greatest successes in cosmology has been the observation of the cosmic microwave background (CMB) and its small fluctuations around a homogeneous background~\cite{Boomerang:2000jdg,WMAP:2012fli,Planck:2018nkj}. These fluctuations are consistent with a nearly scale-invariant primordial spectrum on large scales, while their statistical properties allow us to determine cosmological parameters using ``classical'' cosmological perturbation theory, including the perturbative Boltzmann and Einstein equations. However, standard cosmology does not explain the origin of the perturbations but regards them as initial conditions.

Inflation~\cite{Starobinsky:1980te,Sato:1980yn,Guth:1980zm,Linde:1981mu,Albrecht:1982wi}, a phase of accelerated expansion in the very early Universe, is believed to explain the origin of these fluctuations and thereby connect fundamental quantum theory with our intuitive classical picture of the Universe.\footnote{This Letter considers only quantum fluctuations around a classical background spacetime. A fully quantum-mechanical description including the ``background'' is beyond the scope of this Letter.} Indeed, quantum fluctuations generated during inflation naturally produce a nearly scale-invariant power spectrum, which is regarded as one of the central predictions of inflation consistent with observations~\cite{Planck:2018jri}.

However, this apparent connection between primordial quantum fluctuations and the observed CMB anisotropies conceals a fundamental gap in the foundations of cosmology that has never been justified within quantum theory. In standard late-time cosmology, observations are interpreted on the assumption that one definite c-number realization of primordial perturbations determines the corresponding gravitational potentials, matter perturbations, CMB anisotropies, and ultimately the large-scale structure (LSS) of the Universe. In particular, the photons we observe, together with other matter components, are assumed to have evolved through the same realization. We refer to this as the \textit{one-realization description}, which goes beyond ensemble statistics: the claim that primordial perturbations seeded cosmic structures presupposes a definite history of gravitational and matter evolution. Quantum theory, however, permits a superposition of primordial configurations correlated with distinct histories of gravitational potentials and matter distributions. Invoking wavefunction collapse could select one realization, but such a selection would constitute an additional postulate beyond the unitary evolution of the global quantum state. The fundamental question is therefore how such a quantum state can support a one-realization description while the global superposition persists. This is the cosmological one-realization problem addressed in this Letter.

One concrete manifestation of this problem arises when inferring inflationary parameters from observations: the power spectrum defined by the quantum two-point function, $P_\zeta(k)$, is assumed to correspond to the classical-ensemble power spectrum, $P_{\zeta,\mathrm{ens}}(k)$, used within a one-realization description, schematically expressed as
\begin{align}
    P_\zeta(k)\overset{\mathrm{assumed}}{\simeq}P_{\zeta,\mathrm{ens}}(k).\label{eq:ens_identification}
\end{align}
While such a statistical correspondence is often taken to hold at a formal level, its interpretation in terms of one physically meaningful classical realization requires further justification within quantum theory.

\begin{figure*}[t]
    \centering
    \includegraphics[width=0.9\textwidth]{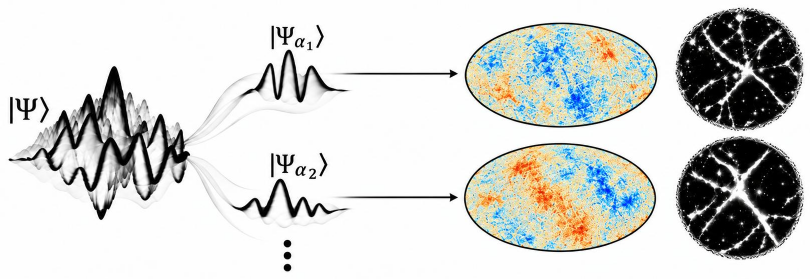}
    \caption{Schematic illustration of the one-realization description of the Universe. We generate this figure using OpenAI's ChatGPT. The global quantum state $|\Psi\rangle$ remains a superposition of different primordial sectors $|\Psi_{\alpha}\rangle$ . When compatible matter records make the conditional matter states associated with different sectors distinguishable, conditioning on a physical record supports a one-realization description at finite resolution, leading to corresponding patterns of CMB anisotropies and large-scale structure. The global quantum superposition persists.
}
    \label{fig:branching}
\end{figure*}

More fundamentally, the one-realization problem calls into question the ontic picture underlying classical cosmology. In classical dynamics, gravitational potentials and matter configurations are assumed to have definite values that exist independently of observation and are shared by all physical systems. Quantum theory, however, does not automatically provide such an absolute, shared classical configuration. Instead, information about a particular configuration can be encoded in the quantum states of individual degrees of freedom or their collections. In this sense, classicality also has an epistemic aspect: what can be regarded as definite depends on the quantum degrees of freedom under consideration and the information contained in their states. We therefore neither assume an ontically definite primordial configuration from the outset nor identify classicality merely with classical statistics. Thus, the intuitive classical picture underlying late-time cosmology and the interpretation of observations must also be reconsidered.\footnote{See also Refs.~\cite{Sudarsky:2009za,Martin:2012pea,Berjon:2020vdv} for relevant issues.}

In this Letter, we address this question by shifting the focus from the quantum state of primordial fluctuations alone to physical \textit{records} of primordial information in late-time matter. Here, records refer to information about primordial curvature perturbations encoded in late-time matter, such that conditional matter states associated with different primordial configurations become distinguishable. In particular, we consider situations in which distinct matter components carry compatible records of the same primordial alternative. Building on the operational perspective on classicality developed in quantum information theory, we show how the quantum-state structure \textit{Cosmological Classical Branches} (CCBs), originally proposed in Ref.~\cite{Kuroda2026CCB}, connects primordial quantum perturbations to a one-realization description of the late-time Universe in the following finite-resolution sense. Figure~\ref{fig:branching} illustrates the schematic picture of CCBs.

We decompose the primordial state into finite-resolution spectral sectors and follow their labels under common unitary evolution. A matter component records a primordial alternative when conditional states associated with different labels are distinguishable, as quantified by root fidelity. When compatible records of the same label exist in disjoint components, a complementary record bounds the interference accessible through operations confined to the other component. Such recorded, locally noninterfering alternatives constitute CCBs. Once CCBs are formed, a one-realization description is thus justified in a branch-conditioned sense, without postulating a globally selected outcome,\footnote{We do not attempt to explain how a single outcome is selected, which is closely related to the so-called wavefunction collapse problem. See, \textit{e.g.},~\cite{Martin:2012pea,Bassi:2012bg}.} while the full quantum state, including other primordial alternatives, remains in superposition.

We illustrate the required distinguishability using an effective two-fluid model involving photons and a second radiation component, whose conditional states encode the same primordial configuration. We further show that the density-based distinguishability grows as modes approach horizon reentry, while it vanishes in the strict soft limit. To demonstrate the latter, we consider the species rest-frame density, whose regular adiabatic response begins at gradient order. This familiar soft behavior identifies the part of the cosmological response that can enter a physical matter record.

For a detailed construction of CCBs, see Ref.~\cite{Kuroda2026CCB}. It compares CCBs with existing related concepts, including squeezing~\cite{Kiefer:1998qe}, decoherence~\cite{Sakagami:1987mp,Brandenberger:1990bx,Calzetta:1995ys}, pointer states~\cite{Zurek:1981xq,Zurek:1982ii,Zurek:2003zz,Kiefer:2006je}, Quantum Darwinism~\cite{Zurek:2009ymk}, spectrum broadcast structure~\cite{Korbicz:2014zed}, and branch structure~\cite{Nelson:2017icq}, highlighting the distinctions and novelty of the CCB framework. In addition, Ref.~\cite{Kuroda2026CCB} further examines photon--dark-matter records and establishes the correspondence between mode averages within a CCB and primordial quantum two-point statistics in the large-volume limit, thereby supporting the assumption in Eq.~\eqref{eq:ens_identification}.

\emph{Cosmological Classical Branches.---}We now explain the criterion for the formation of CCBs~\cite{Kuroda2026CCB}. First, inflationary squeezing identifies the growing-mode curvature perturbation $\zeta_*$ as a natural pointer variable labeling candidate classical evolutions of the Universe, including its late-time evolution. At a reference time $t_*$ when the field eigenstates $|\zeta_*\rangle$ serve as pointer states, we divide the configuration space of the selected primordial curvature modes into finite-resolution cells $\Delta_\alpha$ and introduce the corresponding spectral projectors for the global state $|\Psi_*\rangle$,
\begin{align}
\widehat\Pi_{*,\alpha}=\int_{\Delta_\alpha}\mathcal D\zeta_*\,|\zeta_*\rangle\langle\zeta_*|.\label{eq:primordial-curvature-expansion}
\end{align}
For a pure primordial state, its subsequent evolution can be decomposed as
\begin{align}
|\Psi(t)\rangle=\sum_\alpha\sqrt{p_\alpha}\,|\Psi_\alpha(t)\rangle,\quad|\Psi_\alpha(t)\rangle=\frac{U(t,t_*)\widehat\Pi_{*,\alpha}|\Psi_*\rangle}{\sqrt{p_\alpha}},\label{eq:primordial-state-decomposition}
\end{align}
where $U(t,t_*)$ is the full unitary evolution operator and $p_\alpha=\langle\Psi_*|\widehat\Pi_{*,\alpha}|\Psi_*\rangle$. Each sector $\alpha$ is associated with a representative primordial configuration $\zeta_{*,\alpha}$, which acts as a c-number at the retained resolution. For a general density matrix, the corresponding normalized conditional state is
\begin{align}
 &\widehat\rho_\alpha(t)=\frac{\widehat U(t,t_*)\widehat\Pi_{*,\alpha}\widehat\rho_*\widehat\Pi_{*,\alpha}\widehat U^\dagger(t,t_*)}{p_\alpha},\nonumber\\
&f(\widehat\zeta_*)\widehat\Pi_{*,\alpha}\simeq f(\zeta_{*,\alpha})\widehat\Pi_{*,\alpha},
 \label{eq:sector}
\end{align}
where $p_\alpha=\operatorname{Tr}(\widehat\rho_*\widehat\Pi_{*,\alpha})$. Thus, sector $\alpha$ contains one resolved primordial configuration at the chosen accuracy, while there remains the unresolved fluctuation within the cell, $\delta\widehat\zeta_{*|\alpha}:=\widehat\Pi_{*,\alpha}(\widehat\zeta_*-\zeta_{*,\alpha})\widehat\Pi_{*,\alpha}$.

We now turn to one of the key features distinguishing CCBs from commonly invoked notions of classicality such as decoherence of $\zeta$. After reheating, $\zeta$ is a collective metric--matter variable rather than an additional independent subsystem. CMB observations access information about primordial curvature perturbations through photons. Furthermore, for the formation of LSS, the information should also be shared with the dark matter component as the ``initial conditions'' in standard cosmology. These points are not understood from a quantum state of a single degree of freedom. Our formulation therefore shifts the cosmological classicality problem from asking whether the reduced state of $\zeta$ itself looks classical to asking whether the physical matter in the post-reheating Universe carries \textit{compatible records} of the same primordial realization, as discussed below. In a regime where correlations among different matter components are sufficiently weak, the descendant of the primordial sector $\alpha$ may take the following approximately factorized form:
\begin{align}
    |\Psi_\alpha(t)\rangle \simeq |\gamma_\alpha(t)\rangle \otimes |\nu_\alpha(t)\rangle \otimes |\mathrm{DM}_\alpha(t)\rangle \otimes \cdots.
\end{align}
Thus, the same sector $\alpha$ is interpreted as the global primordial label inherited by all of these conditional matter components and fixes the common sector-dependent gravitational realization entering their dynamics. In practice, we consider coarse-grained mixed states and impose the matching condition for matter components to inherit the same label, as demonstrated in the concrete two-fluid example below.

The spectral decomposition does not by itself imply physically distinguishable matter states for different labels. Hence, the next question is whether this primordial label becomes physically readable through matter, or equivalently, the matter states with different labels are distinguishable. For a physical matter subsystem $F$, we define its conditional reduced state and the root fidelity between different primordial alternatives as
\begin{align}
\widehat\rho_{F|\alpha}(t)&:=\operatorname{Tr}_{\bar F}\widehat\rho_\alpha(t),\label{eq:conditional-state}\\
\mathcal F_F(\alpha,\beta)&:=\left\|\sqrt{\widehat\rho_{F|\alpha}}\sqrt{\widehat\rho_{F|\beta}}\right\|_1,\label{eq:record-fidelity}
\end{align}
where $\bar F$ denotes the complement of $F$ and $\|\cdot\|_1$ is the trace norm. When $\mathcal F_F(\alpha,\beta)\ll1$, the conditional states are distinguishable, and $F$ carries a readable record distinguishing the primordial alternatives $\alpha$ and $\beta$. Thus, the primordial configuration can be inferred from the physical matter state at the specified resolution.

Because the different matter components of interest are expected to respond to the same primordial configuration in a one-realization description, they should yield the same label when read out. Thus, we consider the additional condition that another disjoint matter subsystem $G\subset\bar F$ carries a compatible record of the same primordial label. Here, compatibility means that the conditional states of $F$ and $G$ arise as reductions of the same conditional global state,  rather than representing independently chosen primordial configurations.

Furthermore, sufficiently distinguishable compatible records not only reveal the same primordial label but also suppress locally accessible interference between alternatives, thereby supporting a one-realization description within each sector. Indeed, the complementary record in $G$ bounds the trace norm of the off-diagonal interference part on $F$:
\begin{align}
\| \operatorname{Tr}_{\bar F}|\Psi_\alpha\rangle\langle\Psi_\beta| \|_1
=\mathcal F_{\bar F}(\alpha,\beta)\leq\mathcal F_G(\alpha,\beta).
\label{eq:complementary-record}
\end{align}
This bound, in particular, yields the following inequality for any unitary operation $\widehat V_F$ on $F$:
\begin{align}
\left|\langle\Psi_\beta|(\widehat V_F\otimes\mathbf 1_{\bar F})|\Psi_\alpha\rangle\right|^2
\leq\mathcal F_G(\alpha,\beta)^2.
\label{eq:local-unitary-bound}
\end{align}
This suppression persists under arbitrary trace-preserving quantum operations confined to $F$, even if they modify its record, since the complementary record in $G$ remains untouched. Thus, locally accessible interference is suppressed while global coherence persists. Here, ``local'' refers to subsystem-restricted operations, including apparatus and ancillas, rather than spatial locality.

We can now state the CCB criterion, which requires two disjoint physical matter subsystems $F$ and $G$ to carry compatible records satisfying
\begin{align}
\mathcal F_F(\alpha,\beta)\ll1, \qquad \mathcal F_G(\alpha,\beta)\ll1,
\label{eq:CCB-criterion}
\end{align}
for the resolved alternatives $\alpha\neq\beta$. Further details, including the mixed-state extension and limitations of the local-operation bounds are given in Ref.~\cite{Kuroda2026CCB}.

To clarify the operational meaning of the resulting one-realization description, consider an observer associated with a physical record of $\alpha$. In the ideal orthogonal-record limit, let $\widehat R_\alpha$ project onto this record. The global state conditioned on the record is then
\begin{align}
\widehat\rho_{|R_\alpha}(t):=\frac{\widehat R_\alpha\widehat\rho(t)\widehat R_\alpha}{\operatorname{Tr}[\widehat R_\alpha\widehat\rho(t)]}=\widehat\rho_\alpha(t).
\label{eq:record-conditioned-state}
\end{align}
An observer whose operations are confined to $F$ obtains measurement probabilities given by
\begin{align}
p(m|R_\alpha)=\operatorname{Tr}_F\left[\widehat M_m {\widehat V}_F\widehat\rho_{F|\alpha}\widehat{V}_F^\dagger\right],
\label{eq:record-conditioned-probability}
\end{align}
where $\widehat V_F$ is a unitary operation on $F$ and $\{\widehat M_m\}$ is a positive-operator-valued measure satisfying $\widehat M_m\geq0$ and $\sum_m\widehat M_m=\mathbf 1_F$. Thus, the observer obtains a one-realization description in which the resolved primordial configuration $\zeta_{*,\alpha}$ acts as a common c-number determining the classical metric--matter evolution, while the matter degrees of freedom remain quantum within the sector. This does not imply a globally selected outcome or wavefunction collapse, and the full quantum state retains all other alternatives. Overall, CCBs justify a one-realization description precisely because compatible matter records identify the same primordial configuration while suppressing locally accessible interference between different alternatives.

\begin{figure*}[thb]
 \includegraphics[width=\textwidth]{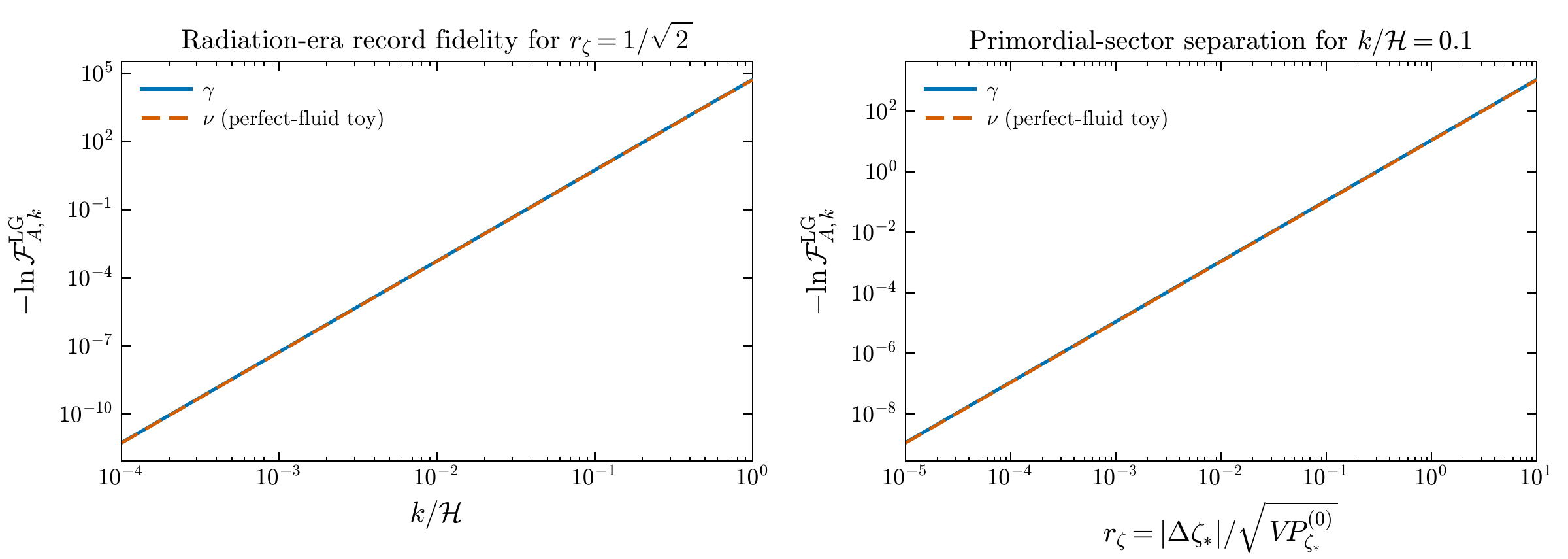}
 \caption{Rest-frame LG fidelity in the two-perfect-radiation-fluid toy model. The figure is reproduced from Ref.~\cite{Kuroda2026CCB}.  The left panel shows the photon and second-radiation-component fidelity exponents from the superhorizon regime through horizon-scale evolution for $r_\zeta(k)=1/\sqrt2$.
 The right panel shows the quadratic dependence on the separation between primordial sectors at $k/\mathcal H=0.1$.  The second radiation component is an ideal perfect-fluid toy component.  The LG states become indistinguishable in the adiabatic soft limit because their rest-frame temperature profiles coincide there.}
 \label{fig:overview}
\end{figure*}

\emph{Two-fluid model including photons.---}We now examine whether ordinary post-reheating matter can form CCBs through compatible records, using the two-fluid model developed in Ref.~\cite{Kuroda2026CCB}. This model considers two perfect radiation fluids, photons $\gamma$ and an idealized neutrino-like component $\nu$, during radiation domination, neglecting anisotropic stress. To characterize their physical responses, we introduce the gauge-invariant rest-frame density perturbation
\begin{align}
\widehat\Delta_{A,\bm k}=\widehat\delta_{A,\bm k}+\frac{4\mathcal H}{k^2}\widehat\theta_{A,\bm k},
\end{align}
where $A=\gamma,\nu$, $\mathcal H=a'/a$ is the conformal Hubble rate, $\widehat\delta_A$ is the fractional density perturbation, and $\widehat\theta_A$ is the velocity divergence. The variable $\Delta_A$ corresponds to the density contrast on the hypersurface comoving with species $A$, and is therefore suitable for characterizing the spatial-locally measurable matter response rather than a gauge-dependent density shift.  In addition, as discussed below, the rest-frame density perturbation is suppressed as $O(k^2/{\cal H}^2)$ in the adiabatic soft limit, so the corresponding density-based records become indistinguishable as $k\to 0$.

As in the case of $\zeta_*$, we consider the density operator within the descendant sector at finite resolution:
\begin{align}
\widehat\Pi_\alpha(t)\widehat\Delta_{A,\bm k}\widehat\Pi_\alpha(t)
&=\Delta_{A,\alpha,\bm k}\widehat\Pi_\alpha(t)+\delta\widehat{\Delta}_{A|\alpha,\bm k},
\end{align}
where $\widehat\Pi_\alpha(t)=\widehat U(t,t_*)\widehat\Pi_{*,\alpha}\widehat U^\dagger(t,t_*)$ and $\delta\widehat{\Delta}_{A|\alpha,\bm k}$ denotes the unresolved fluctuation explained below. Here, the c-number hydrodynamic center is determined by the regular adiabatic transfer,
\begin{align}
\Delta_{A,\alpha,\bm k}=\mathsf T_\Delta^{\rm RD}(k,\eta)\zeta_{*,\alpha,\bm k},
\end{align}
where $\eta$ is conformal time, we have introduced a radiation-domination transfer function $\mathsf T_\Delta^{\rm RD}(k,\eta)=4(\sin y-y\cos y)/y$,  and $y=k\eta/\sqrt{3}$. This is because the linear Heisenberg equations of the density perturbations take the same form as the classical equations of motion, and the adiabatic part associated with the center value $\zeta_{*,\alpha}$ follows the same transfer as the classical solution. On the other hand,  the unresolved fluctuation operator remains within each cell denoted as $\delta\widehat\Delta_{A|\alpha,\bm k}:=\widehat\Pi_\alpha(t)(\widehat\Delta_{A,\bm k}-\Delta_{A,\alpha,\bm k})\widehat\Pi_\alpha(t)$, which can contain the finite-width primordial fluctuation $\delta \zeta_{*|\alpha}$, independent isocurvature perturbations, and microscopic matter fluctuations.

The next question is whether these sector-dependent responses are physically recorded in the quantum states of matter. For this purpose, the conditional radiation states are modeled by local-Gibbs (LG) states~\cite{Hayata:2015lga},
\begin{align}
\widehat\rho_{A|\alpha}^{\rm LG}&=\frac{1}{Z_A[\beta_{A|\alpha}]}\exp\left[-\int_{\Sigma_t}d\Sigma_\mu\,\beta_{A|\alpha,\nu}:\widehat T_A^{\mu\nu}:\right],
\end{align}
where $Z_A$ is the normalization factor, $\widehat T_A^{\mu\nu}$ is the energy--momentum tensor operator, and $u_{A,\alpha}^\mu$ and $T_{A|\alpha}$ are the sector-dependent fluid velocity and local temperature, defining $\beta_{A|\alpha}^\mu={u_{A,\alpha}^\mu}/{T_{A|\alpha}}$. We assume the essential matching condition that the effective LG state reproduces the hydrodynamic center already determined by the Einstein--fluid equations. Specifically, we define the target rest-frame energy-density profile by $\varepsilon_{A,\alpha}^{\rm rf}(x):=\bar\rho_A[1+\Delta_{A,\alpha}(x)]+O(\bar\rho_A\Delta_A^2)$ with the background energy density $\bar\rho_A$ and impose
\begin{align}
u_{A,\alpha,\mu}u_{A,\alpha,\nu}\operatorname{Tr}\left[\widehat\rho_{A|\alpha}^{\rm LG}:\widehat T_A^{\mu\nu}(x):\right]=\varepsilon_{A,\alpha}^{\rm rf}(x).
\end{align}
Together with the fluid velocity fixed by the corresponding adiabatic solution, this condition determines the LG state associated with each primordial sector. For equilibrium radiation, $\bar\rho_A\propto T_A^4$, so the matching yields
\begin{align}
\frac{T_{A|\alpha}-T_A}{T_A}=\frac14\Delta_{A,\alpha}+O(\Delta_A^2).
\end{align}
It should be emphasized that the expectation value appearing in the matching condition does not define or generate the hydrodynamic center: the latter is independently fixed by the classical adiabatic transfer, while the matching selects an effective conditional quantum state whose energy density reproduces that center.

The distinguishability of these conditional states is then quantified by their root fidelity $\mathcal F_A^{\rm LG}(\alpha,\beta):=\left\|
\sqrt{\widehat\rho_{A|\alpha}^{\rm LG}}
\sqrt{\widehat\rho_{A|\beta}^{\rm LG}}
\right\|_1$. In the local-equilibrium approximation, for nearby primordial configurations $\alpha$ and $\beta$, one obtains
\begin{align}
-\ln\mathcal F_A^{\rm LG}(\alpha,\beta)\simeq\frac{\bar\rho_A}{32T_A}\int d^3x\,a^3\left[\Delta_{A,\alpha}-\Delta_{A,\beta}\right]^2,
\end{align}
where $\bar\rho_A$ is the background radiation energy density. Here, we retain the leading local-equilibrium contribution, neglecting higher-order terms in the sector separation and spatial-gradient corrections, which begin at $O[(k_{\rm phys}/T_A)^2]$ for the free radiation model. The calculation also neglects additional distinguishability that could be carried by sector-dependent velocities. In Fourier space, the fidelity can also be computed by considering one nonzero real cosmological Fourier-mode pair $\{\bm k,-\bm k\}$, with the two primordial sectors differing only in this pair, which gives
\begin{align}
-\ln\mathcal F_{A,k}^{\rm LG}(\alpha,\beta)=B_A(k)|\mathsf T_\Delta^{\rm RD}(k,\eta)|^2r_\zeta^2(k),
\end{align}
where
\begin{align}
r_\zeta(k)&:=\frac{|\zeta_{*,\alpha,\bm k}-\zeta_{*,\beta,\bm k}|}{\sqrt{VP_{\zeta_*}(k)}},\quad B_A(k)=\frac{\pi^2}{8}A_s\frac{\bar\rho_A a^3}{T_Ak^3}.
\end{align}
Here $V$ is the comoving normalization volume, $P_{\zeta_*}(k)=2\pi^2A_s/k^3$ is the primordial power spectrum in the scale-invariant approximation, and $A_s$ is its dimensionless amplitude. 

An important consequence is that the fidelity approaches unity in the adiabatic superhorizon limit. Indeed, since $\mathsf T_\Delta^{\rm RD}\simeq(4/9)(k/\mathcal H)^2$, we find, at fixed comoving $k$,
\begin{align}
-\ln\mathcal F_{A,k}^{\rm LG}\simeq\frac{16}{81}B_A(k)r_\zeta^2(k)\left(\frac{k}{\mathcal H}\right)^4+O\left[\left(\frac{k}{\mathcal H}\right)^6\right].
\end{align}
Thus, the radiation states become indistinguishable in the adiabatic soft limit, while their distinguishability can grow toward horizon entry through the amplification of small temperature differences by many microscopic degrees of freedom. This behavior is illustrated in Fig.~\ref{fig:overview} for $B_\gamma=5.59\times10^5$ and an idealized neutrino-like component with $B_\nu/B_\gamma\simeq0.970$.

Importantly, the LG states are effective coarse-grained descriptions of the exact conditional matter states, assumed to arise through sector-independent CPTP channels, so fidelity monotonicity gives $\mathcal F_A(\alpha,\beta)\leq\mathcal F_A^{\rm LG}(\alpha,\beta)$.\footnote{For general discussions of canonical typicality and the emergence of effective thermal states, including descriptions based on quantum channels, see, {\it e.g.}, Refs.~\cite{Goldstein:2005aib,Popescu:2006rhr,Riera:2012ywi,Correia:2023sbx}.} Provided these effective states arise from compatible reductions of the same conditional global state, their small fidelities for $A=\gamma,\nu$ establish microscopic distinguishability and, through the complementary-record bound, local interference suppression, thereby satisfying the CCB criterion. Moreover, observationally normalized parameters give $B_\gamma(k)\simeq6.8\times10^{71}$ at an observational pivot scale $k=0.05\,{\rm Mpc}^{-1}$, indicating potentially enormous many-body amplification even for tiny primordial differences. However, this idealized distinguishability does not directly represent observational accessibility, which can be reduced by limited detector resolution and incomplete access to the matter state. See~\cite{Kuroda2026CCB} for more detailed discussions about coarse-grained descriptions and realistic setups.

Once CCBs are formed by matter records, the relation between statistics inferred from one realization and the quantum power spectrum can be addressed. As shown in Ref.~\cite{Kuroda2026CCB}, the mode-averaged spectrum within a CCB typical with respect to the Born statistics approaches the primordial quantum two-point spectrum when sufficiently many independent Fourier modes contribute in the same way as the conventional arguments~\cite{Peebles1973,Feldman:1993ky,Watts:2002rz}. This provides a quantum-state basis for the statistical correspondence assumed in Eq.~\eqref{eq:ens_identification}. Thus, CCBs provide a quantum-state basis for connecting the predictions of primordial inflationary fluctuations to the classical cosmological history inferred from observations.

\emph{Conclusion.---}In this Letter, we have shown how the one-realization description assumed to connect inflationary scenarios to standard late-time cosmology can be justified within unitary quantum theory through the formation of CCBs. Once CCBs are formed, compatible records in disjoint matter components ensure that the same primordial curvature label can be read out from each component and that locally accessible interference is suppressed. Using a two-radiation-fluid model, we have shown that different primordial alternatives can become distinguishable through many-body amplification in photon states, without requiring wavefunction collapse or a globally selected outcome. Overall, our framework provides an operational quantum-state interpretation of how primordial quantum fluctuations can be recorded in late-time photons and connected to the classical cosmological history inferred from observations.

\begin{acknowledgments}
\emph{Acknowledgments.---}OpenAI's ChatGPT 6 was used as an auxiliary tool for language editing, for creating Figure~\ref{fig:branching}, and for assistance in developing numerical calculation code. The authors take full responsibility for all arguments, calculations, and numerical results presented in this work. %We thank Mohammad Ali Gorji, Amaury Micheli, Swagat Saurav Mishra, Teruaki Suyama, Masahide Yamaguchi, and Yuhang Zhu for their helpful discussions. 
T.K. acknowledges financial support from the Institute for Basic Science (IBS) under the project code IBS-R018-D3 and JST SPRING, Japan Grant Number JPMJSP2180. F.S. acknowledges JSPS Grant-in-Aid for Scientific Research No. 26KJ0060. Y.Y. is supported by IBS under the project code IBS-R018-Y3-2026-a00.
\end{acknowledgments}

\end{document}